\documentclass[a4paper,12pt]{iopart}
\usepackage{graphicx}
\usepackage[T1]{fontenc} 
\usepackage{epstopdf}
\usepackage{bm,amsfonts,lineno,hyperref,amssymb,microtype,float,amsgen,amsbsy}
\usepackage{color}
\usepackage{array}
\usepackage{iopams}
\usepackage{graphicx}
\usepackage{hyperref}
\usepackage{amstext}
\usepackage{savesym}
 \expandafter\let\csname equation*\endcsname\relax
\expandafter\let\csname endequation*\endcsname\relax
\usepackage{amsmath}
\usepackage{txfonts}
\usepackage{microtype}
\usepackage[a4paper, total={6.25in, 8.5in}]{geometry}

\begin{document}

\title{Traversable wormhole on the brane with non-exotic matter: a broader view}

\author{Rikpratik Sengupta$^1$, Shounak Ghosh$^2$, Mehedi Kalam$^1$ and Saibal Ray$^3$}

\address{$^1$ Department of Physics, Aliah University, Kolkata 700160, West Bengal, India}
\address{$^2$ Department of Physics, Indian Institute of Engineering Science and Technology, Shibpur, Howrah 711103, West Bengal, India}
\address{$^3$ Department of Physics, Government College of Engineering and Ceramic Technology, Kolkata 700010, West Bengal, India.}

\ead{$^1$ rikpratik.sengupta@gmail.com}
\ead{$^2$ shounak.rs2015@physics.iiests.ac.in}
\ead{$^1$ kalam@associates.iucaa.in}
\ead{$^3$ saibal@associates.iucaa.in}.
\vspace{10pt}

\begin{abstract}
In this article, the possibility of construction of a traversable wormhole on the Randall-Sundrum braneworld with non-exotic matter employing the Kuchowicz potential has been studied. We have obtained the solution for the shape function of the wormhole and studied its properties along with validity of Null Energy Condition (NEC). The junction conditions at the surface of the wormhole are used to evaluate the model parameters. We also evaluate the surface density and surface pressure for the wormhole. We study the geometrical nature of the wormhole and consider the radial and tangential tidal constraints on a traveller trying to traverse the wormhole. Besides, a linearized stability analysis is performed to obtain the region of stability for the wormhole. Our analysis, besides giving an estimate for the bulk equation of state (EoS) parameter, imposes restrictions on the brane tension, which is a very essential parameter in braneworld physics, and very interestingly the restrictions imposed by our physically plausible and traversable wormhole model are in conformity with those imposed by other braneworld geometries which are not associated with a wormhole solution. Besides, it is important to study such constraints  imposed by geometrical objects such as wormholes on any gravity theory operating at high-energy scales like braneworld, as wormholes are believed to have been formed from massive compact objects of high energy densities. Also, we go on to justify that the possible detection of a wormhole may well indicate that we live on a three-brane universe.
\end{abstract}

\maketitle

\section{INTRODUCTION}
One of the most fascinating and interesting theoretical solutions for the spacetime that can be found from Einstein's
General Relativity (GR) is a wormhole, representing a bridge, i.e., 'throat' like connection between two spacetimes of
the same universe or two different universes. The geometrical structure of wormhole is tubular which spreads and turns
out as asymptotically flat at the infinities and having no singularity or event horizon. The concept of wormhole was
first theorized in 1916 \cite{Ludwig1916}, though the name was not wormhole at the time. In 1935, Einstein and Rosen ~\cite{ER1935} used
the theory of GR to elaborate the idea, suggesting the existence of "bridges" through spacetime, that connects two
isolated areas of spacetime, theoretically creating a shortcut that could reduce travel time and distance, these shortcuts
were termed as Einstein-Rosen bridges, or wormholes. The mathematical structure of wormhole was first obtained by Fuller
and Wheeler~\cite{FH1962} by solving the Einstein Field Equation (EFE) for spherically symmetric spacetime. They have
shown that such solutions were unstable as the throat of the wormhole would have pinched off, due to this any traversing
signal can be trapped in an infinite curvature region, that  makes the wormhole non-traversable. The idea of traversable
wormholes was introduced first by Morris and Thorne~\cite{MT1988}, where they studied the wormhole solutions as a tool for
understanding of GR in a simpler way. According to their prescription for a stable traversable wormhole, the corresponding
energy-momentum tensor must violate the null energy condition (NEC) of GR in such a way that pinching-off of the throat can
be prevented, which is known as the flaring out state of the wormhole. Further, a sustainable wormhole may allow the
possibility of time-travel, allowing closed time-like curves violating causality~\cite{MTY1988}. The infringement of NEC
indicates the existence of exotic matter in the throat of the wormhole. Though such `exotic' matter has not been observed till date,
its presence has been hypothesized in theories for attempting to explain the present cosmic acceleration~\cite{Padmanabhan,Copeland,Caldwell,Frieman}.
Lot of researchers have studied the traversable wormholes admitting exotic matter under the framework of GR which can be found in~\cite{Barcelo,Hayward,Picon,Sushkov,Lobo,Zaslavskii,Chakraborty}. Presence of stable wormhole demands the violation of NEC 
in the spacetime which immediately suggests the presence of exotic matter in it in the context of GR. So,
in GR it is not possible to construct stable wormhole with matter satisfying the NEC. All these factors motivate us to study the
formation of wormhole with  ordinary matter under the framework of braneworld theory of gravity.

Though Einstein's GR is one of the most successful theory to unreveal various mysteries of the universe, but it still requires
some modification to overcome the shortcomings in the theoretical as well as observational aspects. The confirmation of accelerating
expansion of the universe based on some observational evidences as well as the existence of dark matter led to a theoretical challenge
to GR~\cite{Riess1998,Perlmutter1999,Bernardis2000,Hanany2000,Peebles2003,Padmanabhan2003,Clifton2012,Riess2007,Tegmark2004,Amanullah2010,Komatsu2011}.
In order to address these questions various modified theories of gravity have been developed successively with a modification either
in the  matter-energy part  or in the geometrical part of Einstein field equation.  In the context of modified theories of gravity,
one can claim that Einstein's GR fails to provide adequate descriptions at large energy scales and a more generalized action is
required for the gravitational field. It was Hilbert who obtained the field equations using the action principle admitting linear function of the
scalar curvature (or the Ricci Scalar) $R$ in the gravitational Lagrangian. Most of the modified theories have been developed  by
incorporating different functional form of $R$ in the gravitational Lagrangian of the corresponding Einstein-Hilbert action. There
have been number of attempts made to modify GR at higher energy scales~\cite{MaartensLR,BojowaldLR} in order to obtain a set of
non-singular solutions~\cite{Senovilla}, and also at energy scales corresponding to the present epoch of cosmic evolution to accommodate
naturally the current cosmic acceleration without incorporating exotic matter~\cite{Harko,NojiriR}.

On the other hand, a class of modified theories of gravity have been developed considering higher dimensions to overcome the
problems in GR. Kaluza and Klein~\cite{Kaluza1921,Klein1926} first attempted such modification by adopting one compact extra
dimension. Through this modification, they were able to unify the interactions of gravity and electromagnetism. Recently, Randall
and Sundrum gave a concrete form to the idea of braneworld~\cite{Randall1,Randall2} as a higher dimensional approach to
modify GR from a purely geometrical approach~\cite{Shiromizu}. Braneworlds have been used very effectively as a platform to
investigate cosmological~\cite{Binetruy,Maeda,Langlois,Chen,Kiritsis,Campos,Sengupta2,Maartens}, astrophysical~\cite{Wiseman2,Germani,Deruelle,Wiseman,Visser,Creek}
and collapse~\cite{Pal,Dadhich,Bruni,Govender} problems. Due to the non-locality and non-closure properties of the modified
EFE on the brane, the Minimum Geometric Deformation (MGD) approach has been introduced ~\cite{Ovalle2,Ovalle3,Ovalle5,Ovalle6}, which are the
key tools for studying collapse problems on the brane. While the particles and forces, which are operating within the realm
of standard model of particle physics, are believed to be confined to the brane, the gravitational force is let free to traverse
the higher dimensional embedding spacetime known as the bulk. This propagation of gravity has many interesting effects on the
brane or our observed universe. To mention a few, the weakness of the gravitational force compared to the other forces at the
fundamental level can be explained by its simultaneous presence in the higher dimensional embedding space, leaving behind an
effective reduced manifestation on the 3-brane. Also, a possible explanation of the dark matter can be provided by the gravitational
effect of the second brane in case of the two brane RS- model or simply the gravitational effect of the bulk in case of single brane model.

Wormhole solutions in the context of modified gravity without accounting for extra dimensions have been investigated
in~\cite{Bhawal,Bhadra,Eiroa,Bertolami,Moraes,Agnese,He}. In higher dimensional modified gravity approach, wormhole
solutions can be found in~\cite{Dzhunushaliev,Bronnikov,Lobo2,Banerjee,Chakraborty2,Rahaman1,Wang}. The possibility of
existence of wormholes in the central and outer region of galactic halos has been investigated in~\cite{Rahaman2}
and ~\cite{Rahaman3,Kuhfittig,Kalam}, respectively. Here, we attempt to investigate the possibility of construction of a traversable
wormhole on the braneworld, such that the metric potential in the temporal direction is of the Kuchowicz type~\cite{Kuchowicz}.
The motivation behind the choice of this particular function due to it,s well-behaved, analytic and non-singular nature over a finite range of the radial parameter. The Kuchowicz potential has been widely applied for the modelling compact objects~\cite{Ghosh2019,Biswas2019,Biswas2020} to study their properties which provide satisfactory results for those objects. This motivates us to extend the idea of the implementation of this potential function to construction of braneworld wormholes. Moreover, it can be found in literature~\cite{Gao,Fu,KH,HKL} that the traversable wormholes can be constructed from black holes. As black holes form an undisputed class of compact objects, so it would be particularly interesting to study the possibility of construction of a traversable wormhole by using the Kuchowicz potential as the redshift function. The field equations in RS brane gravity include modifications due to higher dimensional contribution from the bulk. So it is very difficult to obtain an exact solution for the wormhole  with both the metric potentials unknown. In fact, this is also the case in many instances when wormhole solutions are obtained in the relativistic context using exotic matter. Again using the Kuchowicz metric potential Ghosh et al.~\cite{Ghosh2019} has obtain a set of acceptable regular solution for gravastars without having the central singularity at the interior. In a similar way with the choice of Kuchowicz potential as redshift function for the construction of wormhole may be done as the throat of the wormhole must not get pinched-off in order to ensure traversability.
 Also, due to lack of experimental
evidence in favour of exotic matter so far, we choose to consider a linear Equation of State (EoS) corresponding to perfect fluid
matter on the brane, describing ordinary non-exotic matter. Such ordinary matter does not violate the NEC but due to the modified
gravity framework we are using, the effective modification is induced in the matter source which may lead to the violation of the
NEC even in the absence of the exotic matter.

 Here, we have tried to construct solution for a static Lorentzian wormhole on the brane which does not evolve with time and hence there is no possibility of expansion of the throat like an Euclidean wormhole. Since the Lorentzian wormhole is static and local object, we have considered the regularity of the redshift function at finite $r$ only. Such similar redshift functions like the Krori-Barua potential have been used to construct static Lorentzian wormhole in literature\cite{IFI}. Moreover, we have not assumed any form for the shape function like many other works and using the Kuchowicz potential as a redshift function we obtain a solution for the shape function on the brane naturally, which satisfies all the criteria laid down by Morris and Thorne to represent a physically traversable wormhole.

It would be extremely satisfactory if the redshift function would have remained finite at infinite distances from the throat and that is why in many works in the relativistic context, the redshift function is taken simply to be a constant or an inversely dependent function of r from the mathematical point of view, but there is no physical justification for choosing such an ad hoc redshift function. Most of the well-known metric potentials for spherically symmetric solutions have this problem of diverging at infinity just like the Kuchowicz function. We have chosen the Kuchowicz potential as the redshift function, from the physical motivation of it’s property of providing regular solution at the interior of gravastar and as it is well-behaved at all points other than $r$ approaching infinity.

Plenty of wormhole solutions have been studied in the last three decades in the General Relativistic context, as they emerge naturally as solutions to the Einstein Field Equations just like black holes or gravitational waves. It is now largely accepted that in order to construct wormholes in the framework of General Relativity, exotic matter is required as it violates the Null Energy Condition (NEC) required to make the wormhole throat stable and prevent it from collapsing. However, such exotic matter has never been observed and in the context of particle physics, the requirement for such “beyond the standard model” exotic matter can be met by the introduction of extra dimensional idea like the braneworld. For eg., the braneworld models can explain the observational effect of dark matter without involving any exotic matter, just by considering the gravitational effect of matter present in the higher dimensional bulk spacetime\cite{SP}. So, we choose to explore the possibility of construction  of a traversable wormhole in a modified gravity context involving the physical idea of higher dimensional  Randall- Sundrum braneworld in the absence of exotic matter.

There may be a possibility of indirect evidence for wormholes like the one being considered by us, from experiments in particle physics looking out for extra dimensions. As we shall see from our analysis, our model can theoretically constrain some of the fundamental parameters involving higher dimensions, which may also be constrained from an experimental point of view in terms of fundamental particle characteristics, if extra spacetime dimensions are indeed detected. 

Recently there have been investigations exploring the possibility of asymptotically flat, traversable wormholes without using exotic phantom matter within the context of GR\cite{Salcedo,Konoplya}. The model obtained by \cite{Salcedo} describes such wormholes which are $Z_2$-symmetric about the throat, supported by a Maxwell and two Dirac fields without any exotic coupling. However, the model is plagued with certain unrealistic physical features like the simultaneous existence of particles and antiparticles at the throat without getting annihilated, besides nonsmooth metric and matter fields. This physical inconsistencies were removed by \cite{Konoplya} in their latest model by not considering the $Z_2$-symmetry relative to the throat and they found that wormholes asymmetric relative to the throat also do not require any exotic phantom matter at the throat to ensure asymptotic flatness or traversability. Moreover in absence of the $Z_2$-symmetry there is no coexistence of particles and antiparticles at the throat and also the supporting metric and matter fields exhibit smoothness. 
	
In the present article we have studied the mathematical model of the wormhole on the brane including the modified EFE on the braneworld,
solution for the wormhole shape function, validity of the NEC on the brane for ordinary perfect fluid matter and obtaining the possible
numerical values of the constants from the junction conditions for plotting the physical parameters of interest pertaining to the wormhole.
We shall then discuss the physical relevance of the obtained solutions and plots for the physical parameters and draw conclusion based on
it in the following section.

\section{Mathematical Model of the Brane Wormhole}

In this Section we will try to construct a detailed mathematical model of the brane wormhole for linear EoS, where one of the
metric potentials is of the Kuchowicz type. We shall start with the modified EFE on the Randall-Sundrum (RS) II braneworld. We briefly discuss the setup for the higher dimensional gravity framework that we have used. The modifications to the standard relativistic field equations for a static wormhole have been presented and then solved analytically to obtain the shape function.

\subsection{Modified EFE on the Brane}
We begin with the general form of the modified EFE on the 3-brane~\cite{Shiromizu} and move on to obtain the set of EFEs for
the case of a spherically symmetric, static metric. The RS II braneworld model considers a single 3-brane embedded in higher
dimensional bulk having one extra dimension which is the compact space $S_1/\bf{Z}_2$, where $\bf{Z}_2$ refers to the orbifold symmetry or
the reflection symmetry.

The general form of the modified EFE on the 3-brane is given as
\begin{equation}
G_{\mu \nu} = - \Lambda g_{\mu \nu} + \kappa^{2} T_{\mu \nu}^{mod},\label{eq1}
\end{equation}
which may be expanded as
\begin{equation}
G_{\mu \nu} = - \Lambda g_{\mu \nu} + \kappa^{2} T_{\mu \nu}+ \kappa_{5}^{4} S_{\mu \nu} - E_{\mu \nu}.\label{eq2}
\end{equation}

The modified stress-energy tensor $T_{\mu \nu}^{mod}$ contains the corrections to the
standard EFE, besides the matter stress-energy tensor $T_{\mu \nu}$ on the brane. Here, the 4-dimensional
constant have been represented by $\kappa^{2}=\frac{\sigma \kappa_{5}^{4}}{6} $, where $\sigma$ denotes the brane tension. In
the high energy regime the brane tension is small and tends to zero for extremely high energies. We
take $\kappa^{2}=8\pi G=1$ on the brane.

The effective cosmological constant on the brane is given by
\begin{equation}
\Lambda=\frac{\kappa_5^2}{2}\left(\Lambda_5+\frac{\kappa_5^2 \sigma^2}{6}\right).\label{eq3}
\end{equation}

In the RS II set-up the bulk is Anti-de Sitter ($AdS_5$), making the bulk cosmological constant
$\Lambda_5$ negative, and it is fine-tuned with the brane tension in such a way as
to give the value of $\Lambda$ to be zero.

The corrections to the EFE are of two types-(i) Local correction ($S_{\mu \nu}$) which comprises
of terms quadratic in the stress-energy tensor on the brane, and (ii) Non-local correction $( E_{\mu \nu})$
which comprises of the projection of the bulk Weyl tensor on the brane, transferring the gravitational
effect from the bulk.

The local correction is given by
\begin{equation} 
S_{\mu \nu} =  \frac{T T_{\mu \nu}}{12}  - \frac{ T_{\mu \alpha} T^{\alpha}_{\nu}}{4}  + \frac{g_{\mu \nu} }{24} \left[ 3 T_{\alpha \beta} T_{\alpha \beta} - (T^{\alpha}_{\alpha})^{2} \right], \label{eq4}
\end{equation}
where $T = T^{\alpha}_{\alpha}$ and we consider matter on the brane with the perfect fluid stress-energy tensor
\begin{equation} 
 T_{\mu\nu}=\rho u_{\mu} u_{\nu}+ ph_{\mu\nu}. \label{eq5}
\end{equation}

Here $u_{\mu}$ is the 4-velocity, $h_{\mu\nu}=g_{\mu\nu}+u_{\mu} u_{\nu}$ is the induced metric on the
brane, $\rho$ and $p$ are the energy density and pressure,respectively.

The non-local correction has the form
\begin{equation}
{E}_{\mu \nu} =  C^{(5)}_{ACBD} n^{C} n^{D} g_{\mu}^{A} g_{\nu}^{B}=-\frac{6}{\sigma}\left[Uu_{\mu} u_{\nu}+Pr_{\mu}r_{\nu}+h_{\mu\nu}\left(\frac{U-P}{3}\right)\right],\label{eq6}
\end{equation}
where $C^{(5)}_{ACBD}$ is the bulk Weyl tensor, $n^C$ is unit normal spacelike vector, $r_{\mu}$ denotes
projected radial vector, $U$ and $P$ denote the bulk energy density and bulk pressure respectively.
${E}_{\mu \nu}$ is symmetric and traceless. It has no components orthogonal to the brane.

 For a static, spherically symmetric matter distribution on the 3-brane, the line element is given by
\begin{equation}\label{eq7}
ds^2=-e^{\nu(r)}dt^2+e^{\lambda(r)}dr^2+r^2(d\theta^2+sin^2\theta d\phi^2).
\end{equation}

The modified EFE on the brane, given by Eq. (\ref{eq2}) can be computed as
\begin{eqnarray}
e^{-\lambda}\left(\frac{\lambda^\prime}{r}-\frac{1}{r^2}\right)+\frac{1}{r^2}
=\left(\rho  \left( 1+\frac {\rho }{2 \sigma} \right) +{\frac {6 U}{\sigma}}\right),\label{eq8}\\
e^{-\lambda}\left(\frac{\nu^\prime}{r}+\frac{1}{r^2}\right) -\frac{1}{r^2}
=\left(p +{\frac {\rho  \left( p +\frac{\rho}{2} \right)}
{\sigma}}+{\frac {2U}{\sigma}}+{\frac {4 P}{\sigma}}\right),\label{eq9}\\
 e^{-\lambda}\left(\frac{\nu''}{2}-\frac{\lambda^\prime \nu^\prime}{4}+\frac{{\nu^\prime}^2}{4}+\frac{\nu^\prime-\lambda^\prime}{2r}\right) = \Bigg(p+{\frac{\rho \left(p+\frac{\rho}{2}\right)}{\sigma}}+{\frac{2U}{\sigma}}-{\frac {{2 P}}{\sigma}}\Bigg).\label{eq10}
 \end{eqnarray}

 The brane energy-momentum tensor and the overall effective energy momentum tensor are both conserved
 separately. So, the conservation equation on the brane reads the same as in GR
 \begin{equation} \label{eq11}
 \frac{dp}{dr}=-\frac{1}{2}\frac{d\nu}{dr}(p+\rho).
 \end{equation}

 \subsection{Solution to the Wormhole Shape Function}

 The line element for a static, spherically symmetric wormhole can be written as
 \begin{equation}
 ds^2=-e^{\nu(r)}dt^2+\frac{dr^2}{1-\frac{b(r)}{r}}+r^2(d\theta^2+sin^2\theta d\phi^2), \label{eq12}
 \end{equation}
 where $b(r)$ denotes the shape function of the wormhole.

 Rewriting the above field equations (\ref{eq8})-(\ref{eq10}) for a wormhole with static and spherically symmetric
 matter distribution, in terms of the shape function $b(r)$ we get eventually
\begin{eqnarray}
 \frac{b^\prime}{r^2}=\rho \left(1+\frac {\rho}{2\sigma}\right) +{\frac{6(X\rho+Y)}{\sigma}}, \label{eq13}\\
 \left(1-\frac{b}{r}\right)\left(\frac{\nu^\prime}{r}+\frac{1}{r^2}\right) -\frac{1}{r^2}
 =p +{\frac {\rho\left(2 p +\rho \right)} {2\sigma}}+{\frac {2(X\rho+Y)}{\sigma}}+{\frac {4 \omega(X\rho+Y)}{\sigma}},\label{eq14}\\
 \left(1-\frac{b}{r}\right)\left(\nu'' + {{\nu^\prime}^2} +\frac{\nu^\prime}{r}  \right) -\frac{b^\prime -b}{2r}\left({\nu^\prime} +\frac{1}{r} \right)
 =p+{\frac{\rho \left(2 p +\rho \right)}{2 \sigma}}+{\frac {2(X\rho+Y)}{\sigma}}-{\frac {2 \omega(X\rho+Y)}{\sigma}}. \label{eq15}
 \end{eqnarray}

 It can be observed from the Eqs. (\ref{eq8})-(\ref{eq11}) that there are six unknowns parameter and we have only four equations on the brane. 
 In order to solve these equations we have followed the previous prescriptions in the
 brane gravity framework and considered the bulk EoS as $P=\omega U$~\cite{Sengupta,Castro}, where $\omega$ is (EoS) parameter, and the energy density of the
 brane and bulk are related by $U=X\rho+Y$~\cite{Sengupta}, where $X$ and $Y$ are the constants and determined from the boundary conditions.

 Now, we consider the metric potential $e^{\nu(r)}$ to be of the Kuchowicz type~\cite{Kuchowicz} which is given as
 \begin{equation}
 e^{\nu(r)}=e^{Br^2+2\ln C} ,\label{eq16}
 \end{equation}
 where $B$ and $C$ are arbitrary constants. The dimension of $B$ is $[L^{-2}]$ and $C$ is a dimensionless constant. The redshift function diverges as $r$ tends to infinity. We are considering a model here for a static wormhole and as there is no time evolution or expansion of the wormhole throat, so the finiteness of the redshift function is physically significant in the vicinity of the wormhole. It can be seen that for all finite values of $r$, the redshift function exhibits regularity and is capable of producing a well behaved and stable, traversable wormhole with the obtained shape function satisfying all the criteria as per the Morris-Thorne prescription, as discussed in details in the concluding section.

As argued earlier, due to lack of observational confirmation of exotic matter which is usually preferred for constructing
traversable wormholes, we chose to propose an ansatz for the EoS describing ordinary non-exotic matter on the brane as follows
\begin{equation} \label{eq17}
p(r)=\mu \rho(r),
\end{equation}
where $\mu>0$.

Using Eqs. (\ref{eq11}), (\ref{eq16}) and (\ref{eq17}) we get
\begin{equation} \label{eq18}
\rho(r)=C_1e^{-\frac{(\mu+1) Br^2}{2\mu}},
\end{equation}
where $C_1$ is the integration constant which can be determined from the matching condition. The variation
of the energy density $\rho(r)$ on the brane along the radial distance $r$ is plotted in Fig. 1. The figure
clearly indicates that the matter density at the throat is much higher and gradually decreasing with respect
to the radial parameter attains a minimum value at the boundary.

\begin{figure*}[thbp] \label{rho}
	\centering
 	\includegraphics[width=0.5\textwidth]{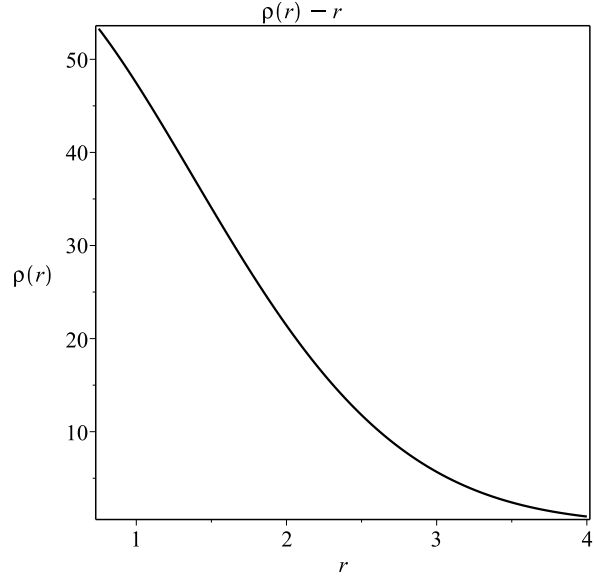}
 	\caption{Variation of the density with respect to $r$.}
 \end{figure*}

 The obtained $\rho$ can be plugged in along with Eq.~(\ref{eq16}) in the field Eq.~(\ref{eq15}) to obtain the shape function for the wormhole given by

 \begin{eqnarray}
 b(r) =\frac{2r e^{-2Br^2}}{Bk^4\sigma }\left(-\frac{ \mu{C_1}(\sigma \mu k^4-2 X (\omega-1)){e^{\frac{B r^2(3\mu-1)}{2\mu}}}}{(3\mu-1)(2Br^2+1)}+
   \frac{k^4}{(\mu-1)(2Br^2+1)}\left(-\frac{{C_1}^2(2\mu+1) \mu{ e^{{\frac{Br^2(\mu-1)}{\mu}}}}}{4}\right. \right.\nonumber\\
 \left.\left.+\frac{\sigma (\mu-1)}{2} \left( B (2 B r^2+1)+  Y (\omega-1) \right)e^{2Br^2}+C_2 B \right) \right), \label{eq19}
 \end{eqnarray}

 where $C_2$ is the integration constant.
\begin{figure*}[thbp] \label{SF}
	\centering
	\includegraphics[width=0.4\textwidth]{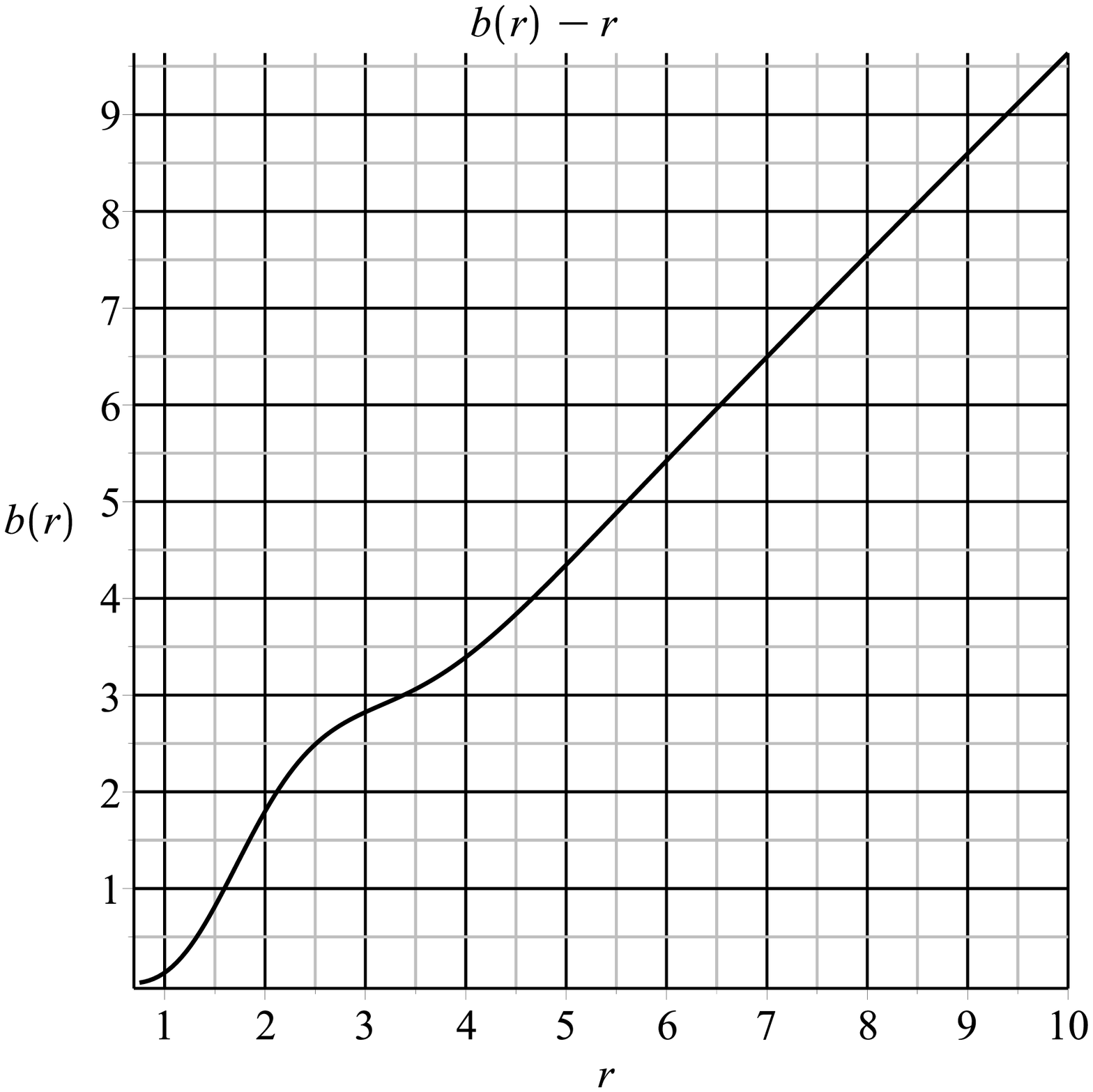}
	\includegraphics[width=0.4\textwidth]{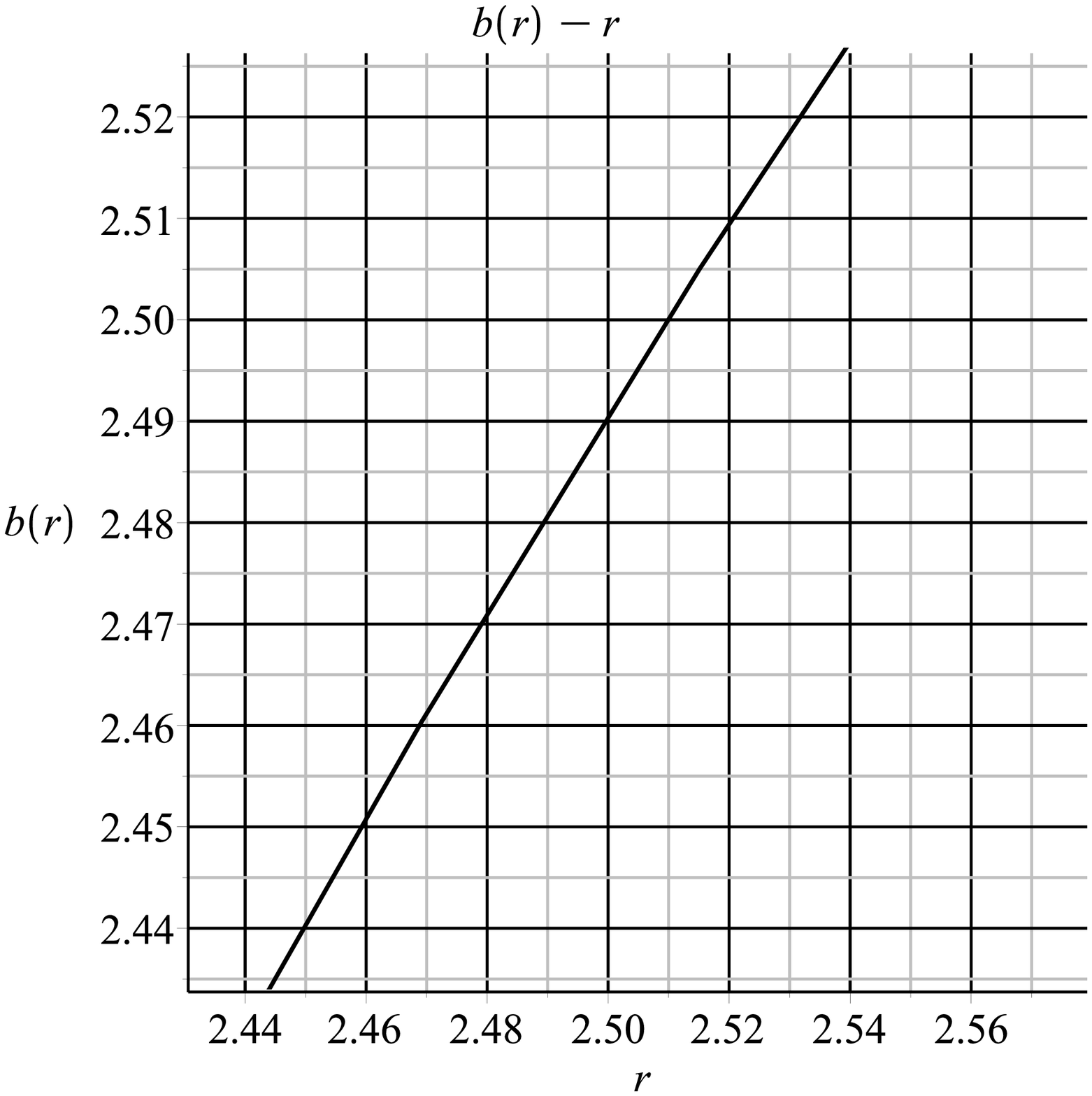}
	\caption{(Left Panel)Variation of the shape function with respect to $r$;        (Right Panel)Variation of the shape function with respect to $r$ in vicinity of $r=2.5$}
\end{figure*}

The shape function is found to be dependent on all the brane and bulk parameters  including the brane tension. The variation of the shape function $b(r)$ along the radial distance $r$ has been plotted in Fig. 2. Apparently, in the vicinity or $r=2.5$ $b(r)=r$,but on close inspection by enlarging the plot in the above mentioned range it can be seen that $b(r)<r$ (right panel of Fig. 2). 

The obtained shape function is found to satisfy all the criteria to represent a traversable wormhole, without the requirement for exotic matter violating the null energy condition (NEC). The physical justification for this comes from the fact that, as a result of the higher dimensional corrections to 4-D gravity, there is an “effective matter” description on the brane, which results in the violation of the NEC, as we shall find in the following susection.

The properties of the shape function that can be inferred from the plot and its feasibility in construction of wormhole has been
 discussed later in the concluding section.

 \subsection{Validity of NEC}

 One of the most important energy conditions appearing in GR is the NEC  given by  $T_{\mu \nu}k^{\mu}k^{\nu} \geq 0 $,
 where $k^{\mu}$ denotes a null vector. This  reduces to $\rho+p \geq 0 $ for a perfect fluid. So, for the ordinary matter we have considered
 on the brane having EoS of form given by Eq. (\ref{eq17}), the NEC holds good. However, as we are concerned with the effective stress-energy
 tensor on the brane due to the local and non-local  correction terms, so it is the effective energy density and pressure that we must consider
 for constructing the NEC on the brane. Hence, the NEC becomes $\rho^{eff}(r)+p_{r}^{eff}(r) \geq 0 $, where it is to be
 noted from Eqs. (\ref{eq9}) and (\ref{eq10}) that the effective pressures in the radial and transverse directions differ due to difference
 in contribution from the bulk pressure term.

 Here
 \begin{equation}
  \rho^{eff}=\left[\rho  \left( 1+\frac {\rho }{2 \sigma} \right) +{\frac {6 U}{\sigma}}\right], \label{eq20}
 \end{equation}
 and
 \begin{equation}
 p_{r}^{eff}=\left(p  +{\frac {\rho  \left( p +\frac{\rho}{2} \right)}
 	{\sigma}}+{\frac {2U}{\sigma}}+{\frac {4 P}{\sigma}}\right). \label{eq21}
 \end{equation}

 The sum of the effective energy density and effective radial pressure can be computed to be
 \begin{eqnarray}
 &&\rho_{eff}+p_{reff}\nonumber\\
 &=&\frac{1}{k^4\sigma} \left(k^4{C_1}^2 (\mu+1)e^{-\frac{(\mu+1) Br^2}{\mu}}+4C_1 \left(\frac{\sigma(\mu+1) k^4}{4} +X(\omega+2)\right) e^{-\frac{(\mu+1) B r^2}{2\mu}}+4Y (\omega+2)  \right). \label{eq22}
 \end{eqnarray}

 \begin{figure*}[thbp] \label{NEC}
 \centering
 	\includegraphics[width=0.5\textwidth]{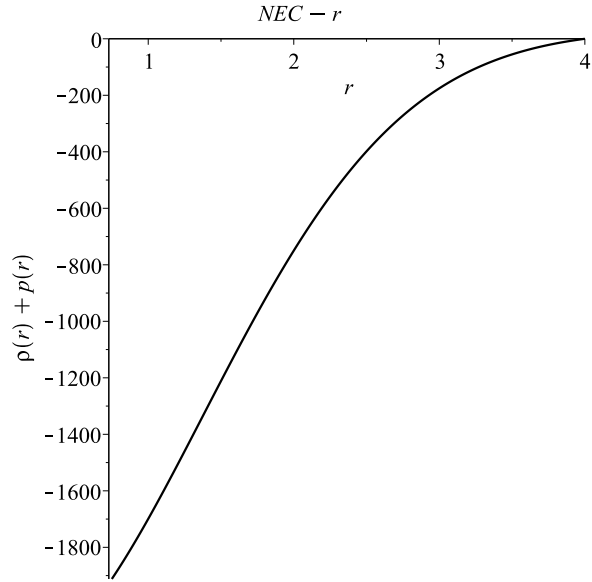}
 	\caption{Variation of the NEC with respect to $r$.}
 \end{figure*}

 The variation of $\rho_{eff}+p_{r}^{eff}$ along $r$ has been plotted in Fig. 3. We find that
 the NEC is violated for effective matter distribution on the brane.

We have not used exotic matter to violate the Null Energy Condition (NEC). The higher dimensional contributions in Equation 2 appear as an effective modification to the stress energy tensor on the brane, but we have not introduced any such matter physically. This is the benefit of using the RS brane gravity to construct a traversable wormhole as compared to General Relativity, as we do not have to introduce any additional exotic matter which has not been detected, but rather the NEC violation is enforced from the higher dimensional contribution itself.

 \subsection{The Junction Conditions}
The exterior spacetime is comprises of vacuum, the line element for which may be described by the Schwarzschild solution
on the brane\cite{DM}. As shown by \cite{DM} in their seminal work, the static, spherically symmetric vacuum solution on
the brane is modified from GR by the effect of a "tidal charge" which is produced a result of the projection of the higher
dimensional bulk gravity effects onto the 3-brane. This leads to the effective brane Schwarzschild solution behaving as
Reissner-Nordstr$\ddot{o}$m (RN) solution of GR even in the absence of any electric charge. This is a very interesting consequence
of single brane RS gravity. So the line element for the spacetime exterior metric to the
wormhole is written as
\begin{equation}\label{eq23}
ds^2=\left(1-\frac{2M}{r}+\frac{Q}{2r^2}\right)dt^2-\left(1-\frac{2M}{r}+\frac{Q}{2r^2}\right)^{-1}dr^2-r^2(d\theta^2+\sin^2 \theta d\phi^2),
\end{equation}
where $M$ is the total mass of the wormhole and $Q$ is the tidal charge. The tidal charge parameter can take both positive and
negative values. The value may be quite large in very strong gravity regimes but even a small value will ensure significant
higher dimensional correction to GR.

Presence of matter on the surface of wormhole must produce an extrinsic discontinuity which generates an intrinsic surface energy
density and surface pressure. The surface behaves as the junction between the spacetime inside the wormhole and its exterior, which
makes the wormhole a geodesically complete manifold with ordinary matter characterizing the configuration of the wormhole. Thus,
according to the fundamentals of junction conditions there should be a smooth matching between the two spacetimes at the junction.
Again, the continuity of the metric coefficients at the junction surface do not guarantee the continuity of its derivatives. So, in
order to determine the surface stress-energy $S_{ij}$ one can use formalism suggested by Darmois-Israel~\cite{Darmois,Israel}.

Now intrinsic surface stress energy tensor $S_{i}^{j}$ can be obtained from Lanczos equation~\cite{Lanczos1924,Sen1924,Perry1992,Musgrave1996}
in the following form
\begin{equation}\label{eq24} S_{j}^{i}=-\frac{1}{8\pi} (\kappa_{j}^{i}-\delta_{j}^{i} \kappa_{k}^{k}),
\end{equation}
the discontinuity in the second fundamental form can be written as
\begin{equation}\label{eq25}
\kappa_{ij}=\kappa_{ij}^{+}-\kappa_{ij}^{-},
\end{equation}
whereas the second fundamental form is given by
\begin{equation}\label{eq26}
\kappa_{ij}^{\pm}=-n_{\nu}^{\pm}\left[\frac{\partial^{2}X_{\nu}}{\partial \xi^{i}\partial\xi^{j}}+
\Gamma_{\alpha\beta}^{\nu}\frac{\partial X^{\alpha}}{\partial \xi^{i}}\frac{\partial X^{\beta}}{\partial
\xi^{j}} \right]|_S,
\end{equation}
where the unit normal vector $n_{\nu}^{\pm}$ are defined as
\begin{equation}\label{eq27}
n_{\nu}^{\pm}=\pm\left|g^{\alpha\beta}\frac{\partial f}{\partial X^{\alpha}}\frac{\partial f}{\partial X^{\beta}}
\right|^{-\frac{1}{2}}\frac{\partial f}{\partial X^{\nu}},
\end{equation}
with $n^{\nu}n_{\nu}=1$. Here $\xi^{i}$ is the intrinsic coordinate of the surface of the wormhole and $f(x^{\alpha}(\xi^{i}))=0$
is the parametric equation of it. Here $+$ and $-$ corresponds to exterior and interior spacetime respectively  of the wormhole.

Now the surface stress energy tensor, for the spherically symmetric spacetime, can be written as $S_{i}^{j}= diag(-\Sigma,\mathcal{P})$.
Where $\Sigma$ and $\mathcal{P}$ are the surface energy density and surface pressure respectively, which can be obtained at the surface
of the wormhole (i.e at $r=R$)  by the following equations
\begin{eqnarray}\label{28}
\Sigma  =-\frac{1}{4\pi R}\bigg[\sqrt{e^\lambda}\bigg]_-^+\nonumber\\
=\frac{1}{4\pi R}\left[\sqrt{\left(1-\frac{2M}{R}+\frac{Q}{2r^2}\right)} - \sqrt{1-\frac{2 e^{-2 BR^2}}{Bk^4\sigma}\left(H_1-\frac{\mu C_1 (\sigma\mu k^4-2 X ( \omega-1)){e^{\frac{BR^2 (3\mu-1) }{2\mu}}}}{{(3\mu-1)(2B R^2+1)}}\right)} \right],\\
\mathcal{P}  =\frac{1}{16\pi R } \bigg[\bigg(\frac{2f+f^\prime R}{\sqrt{f}}\bigg) \bigg]_-^+ =\frac{1}{16\pi R}\left[\frac{2(1-\frac{M}{R})}{ \sqrt{1-\frac{2M}{R}+\frac{Q}{2r^2}}}-\frac{F1}{F2}\right],
\end{eqnarray}
where

\begin{eqnarray}
H_1= -\frac{{C_1}^2 (2\mu+1) k^4\mu {e^{{\frac{BR^2 (\mu-1)}{\mu}}}}}{4(\mu-1)(g_3-1)}+ \frac{(4B^2k^4R^2\sigma+Y(\omega-1)  ){ e^{2 BR^2}}+C_2 B k^4\sigma }{4(2g_3-1)},\nonumber\\
\\
F1=\frac{2C_1 g_1(2X (\omega-1)-\sigma \mu k^4) (\mu-1)\left(g_2(2g_3-1)-2\mu\right)  }{k^4\sigma ( 3\mu-1 )(\mu-1)(2g_3-1 )^{2}B}  +\left(-3 k^4{C_1}^2(2\mu+1) \right.\nonumber\\
\left.\left.\left(\left(2g_2 g_3-\mu\right)\right)g_1+2(\mu-1)\left(C_2 B k^4\sigma(2BR^2 g_3-1){e^{-2BR^2}}-Y(\omega-1)\right)\right)(3\mu-1)\right), \nonumber\\
\\
F2=\sqrt{1-\frac{2 e^{-2 BR^2}}{Bk^4\sigma(2B R^2+1)}\left(-\frac{\mu C_1 (\sigma\mu k^4-2 X ( \omega-1)){e^{\frac{BR^2 (3\mu-1) }{2\mu}}}}{{(3\mu-1)}}+H_1\right)}.\nonumber\\
\end{eqnarray}
Here, $g_1={e^{-{\frac{BR^2(\mu+1)}{\mu}}}}$,  $g_2=BR^2 (\mu+1)$ and $g_3=(BR^2+1)$.

Now for the static wormhole the surface pressure as well as the surface energy must vanish at the boundary surface. So we have $\sigma =\mathcal{P}=0$ at the boundary, which eventually provides the condition
\begin{equation}\label{eq30}
 b(r)|_{r=R}=2M+\frac{Q}{2R^2}.
\end{equation}

In order to calculate the different unknown constants we have used the following boundary conditions:\\

(i) From the junction condition at the boundary (from Eq.(\ref{eq30})) $b(r)|_{r=R}=2M+\frac{Q}{2R^2}$.\\

(ii) Continuity of the metric potential $g_{tt}$ and its derivative $\frac{\delta g_{tt}}{\delta r}$ at the surface boundary, i.e., at $r=R$.\\

Now we choose appropriate realistic values for the physical parameters namely $\ \sigma = 10^3 \ Mev/ Fermi^3, Q=0.00018,  M=1.7\ M_\odot,\  r_0=0.75,\  R=4 \ km $, where $r_0$
is the throat radius (in $km$). Making use of the above conditions, we obtain the values of different parameters
associated with the wormhole as: $X = -1943.448951, Y = 1684.121721, C_1 = 64.85766573, C_2 = -61.40896649,\  \omega = 0.9157485630 $.
These values of the parameters have been used to plot Figs.~(\ref{rho})-(\ref{NEC}). All the plots
have been started from the throat (i.e. $r=r_0$) upto the surface boundary (i.e at $r=R$).

We have not adopted any ad-hoc values for the free model parameters. We explicitly obtain their values from the Israel-Darmois junction conditions. Only the essential physical parameters like realistic physical value of brane tension for the energies corresponding to the present universe and tidal charge have been taken from other works\cite{Sengupta}. There is no constraint on upper limit of wormhole mass for our model, as we have not used any unobserved exotic matter for wormhole construction. So, we have chosen a realistic value.

\subsection{Tidal acceleration}

In order to ensure the traversability of our wormhole, it has to be accounted that tidal forces do not become so large that the observer travelling through the wormhole gets ripped apart. This implies that the tidal acceleration experienced by the traveller should not exceed a realistic value, like the acceleration due to gravity on the earth \cite{MT1988}. The tidal acceleration has both radial and tangential components which must be individually constrained to be less than the acceleration due to gravity on the earth. These two radial and tangential tidal constraints may be expressed as inequalities in terms of the Riemann curvature tensor of the form

\begin{equation}
	|R_{rtrt}|=|(1-\frac{b}{r})\big[\frac{\nu"}{2}+\frac{\nu'^2}{4}-\frac{b'r-b}{2r(r-b)}.\frac{\nu'}{2}\big]|\leq g_{earth}.
\end{equation}

\begin{equation}
	\gamma^2 |R_{\theta t \theta t}|+\gamma^2v^2 |R_{\theta r \theta r}|=|\frac{\gamma^2}{2r^2}\big[v^2(b'-\frac{b}{r})+(r-b)\nu'\big]|\leq g_{earth}.
\end{equation}

where $\gamma=\frac{1}{\sqrt{1-v^2}}$ and $v$ is the traveller’s velocity. If we consider realistically, velocity of the traveller $v\ll 1$ which implies $\gamma\approx 1$ and substituting the form of $b(r)$ given in Eq. (19), one can get at the throat of the wormhole
\begin{equation}\label{99}
v \leq 0.01207442556\sqrt{g_{earth}}.
\end{equation}

The velocity of the traveller is restricted by the above inequality if the tangential tidal acceleration is to be kept sufficiently
small such that the wormhole remains traversable. We also find that the radial tidal acceleration is sufficiently small enough at the
throat to allow traveller to pass through without getting ripped apart. Thus, the tidal acceleration as obtained above, also ensures the traversability of our proposed wormhole model.

\subsection{Linearized stability analysis}

For performing qualitative stability analysis, we consider the radius of the wormhole throat to be a function of proper time. Let us denote the throat radius as $r_0=x(\tau)$. The energy density turns out to be
\begin{equation}
	\sigma=-\frac{1}{2\pi x}\sqrt{f(x)+\dot{x}^2},
\end{equation}
and the pressure is given by
\begin{equation}
	p=\frac{1}{8\pi}\frac{f'(x)}{\sqrt{f(x)}}-\frac{\sigma}{2}.
\end{equation}

Here $f(x)=1-\frac{2M}{x}+\frac{Q}{2x^2}$, where $M$ is the mass of the wormhole and $Q$ is the tidal charge which arises due to higher dimensional effects.

From the conservation equation we get an equation of motion of the form
\begin{equation}
	\dot{x}^2+V(x)=0,
\end{equation}
where the potential $V(x)$ is given by
\begin{equation}\label{LS2}
	V(x)=f(x)-[2\pi x \sigma (x)]^2.
\end{equation}

It is now possible to perform stability analysis of the wormhole by considering a linearization around $x_0$, which we assume to be a static solution to the equation of motion Eq.(41).

\begin{figure*}[thbp] \label{NEC}
	\centering
	\includegraphics[width=0.5\textwidth]{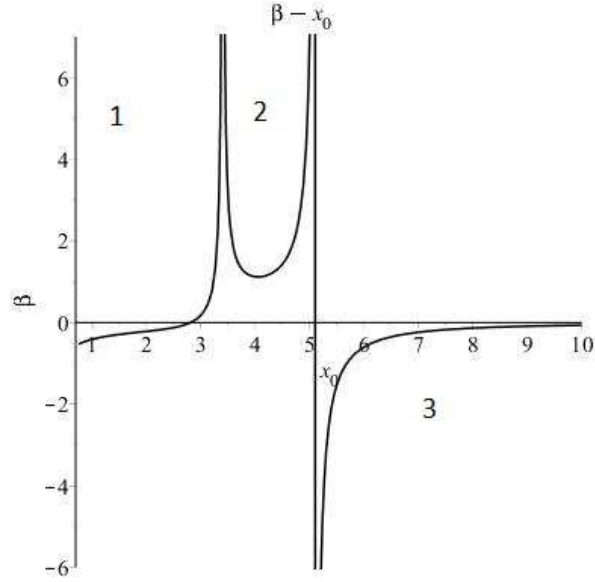}
	\caption{Plot of $\beta$ vs $x_0$.}
\end{figure*}

We consider Taylor series expansion of the potential around the assumed static solution upto second order, which gives
\begin{equation}\label{LS3}
V(x)=V(x_0)-V'(x_0)(x-x_0)+\frac{1}{2}V"(x_0)(x-x_0)^2+O[(x-x_0)^3],
\end{equation}
where prime denotes derivative with respect to x.

As we have considered a static wormhole spacetime in Eq.(12),$V(x_0) = 0$ and
$V'(x_0) = 0$.  The wormhole will be stable if and only if $V"(x_0) > 0$. We define the parameter $\beta$
\begin{equation}\label{LS4}
\beta(\sigma)=\frac{\delta p}{\delta \sigma}.
\end{equation}

It turns out that
\begin{equation}\label{LS5}
V''(x)=f''(x)-8 \pi^2 [ (\sigma +2p)^2+ \sigma (\sigma+p)(1+2\beta)
\end{equation}

Therefore, for the condition we obtain for the wormhole to be stable is
\begin{equation}\label{LS7}
\beta< \frac{\frac{f''(x_0)}{8\pi^2}-(\sigma +2p)^2-2\sigma(\sigma+p)}{4 \sigma (\sigma +p)},
\end{equation}
which on substituting the values of $\sigma$ and $p$ can be written as
\begin{equation}\label{LS8}
\beta< \frac{x_0^2 (f_0')^2-2x_0^2 f_0'' f_0}{4 f_0(a_0 f_0' -2f_0)}-\frac{1}{2}
\end{equation}

We have obtained the stable regions denoted by regions 1,2 and 3 in Fig. (4). So, our model is found to be stable by performing a linearized stability analysis.

\subsection{Acceleration and nature of the wormhole}

An interesting feature of the wormhole can be understood from the redshift function. Its derivative with respect to the radial coordinate determines whether the wormhole has an "attractive" or "repulsive" geometrical nature. The 4-velocity of an observer at rest is given by
\begin{equation}
	U^{\mu}=\frac{dx^{\mu}}{d\tau}=(e^{-\frac{\nu(r)}{2}},0,0,0),
\end{equation}
where $\tau$ denotes proper time. The 4-acceleration of the static observer is given by $a^{\mu}=U^{\mu}_{;\nu}U^{\nu}$, which has the radial component given by
\begin{equation}
	a^r=\frac{\nu'}{2}\bigg(1-\frac{b(r)}{r}\bigg).
\end{equation}

The geodesic equation gives the equation of motion for an initially static test particle, moving in the radial direction as
\begin{equation}
	\frac{d^2 r}{dt^{\tau}}=- \Gamma^r_{tt}\bigg(\frac{dt}{d\tau}\bigg)^2= -a^r
\end{equation}

The wormhole geometry is “attractive” if $a^r > 0$, which implies that an observer should have an outward-directed radial acceleration to prevent being pulled into the wormhole. Alternatively, if it turns out that $a^r < 0$, then it implies that the wormhole geometry is “repulsive” meaning that an observer must have an inward-directed radial acceleration in odrer to prevent oneself from being pushed away from the wormhole.
\begin{eqnarray}	
a^r=\frac {1}{B \left( \mu+1 \right)^{2}{k}^{4}\sigma} \left( 8Br\pi\mu{ C_1} \left( \mu+1 \right)  \left( {k}^{4}\sigma+6X
 \right) {{\rm e}^{-\frac{1}{2}{\frac { \left( \mu+1 \right) B{r}^{2}}{\mu}}}}-4\sqrt {2}{\pi }^{\frac{3}{2}}\sqrt {{\frac { \left( \mu+1 \right) B}{\mu
}}}{\mu}^{2}{ C_1} \left( {k}^{4}\sigma+6X \right) \right. \nonumber\\
\left.{\rm erf}\left(\frac{1}{2}\sqrt {2}\sqrt {{\frac { \left( \mu+1 \right) B}{\mu}}}r
\right)+2 \left( \mu+1 \right) B\mu\pi {{\rm e}^{-{\frac { \left( \mu+1 \right) B{r}^{2}}{\mu}}}}{{ C_1}}^{2}{k}^{4}r-\sqrt
{{\frac { \left( \mu+1 \right) B}{\mu}}}{\mu}^{2}{\pi }^{\frac{3}{2}}{\rm erf} \left(\sqrt {{\frac { \left( \mu+1 \right) B}{\mu}}}r\right){{
C_1}}^{2}{k}^{4}\right..\nonumber\\
\left.+{B}^{2} \left( \mu+1 \right) ^{2} \left( \sigma \left( r-{ \_C2} \right) {k}^{4}-16\pi Y{r}^{3} \right) \right)
\end{eqnarray}
\begin{figure*}[thbp] \label{NEC}
	\centering
	\includegraphics[width=0.5\textwidth]{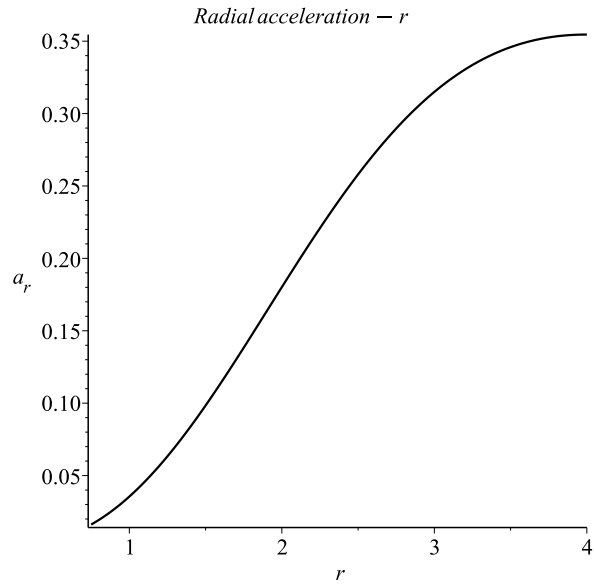}
	\caption{Plot of radial component of acceleration with respect to $r$.}
\end{figure*}
As we can see from Fig. 5 for our wormhole model on the RS braneworld, it turns out that the radial component of 4-acceleration is positive, which indicates an attractive wormhole geometry.

\vspace{0.5cm}

As, we shall see in the following section, our model also interestingly reproduces some theoretical constraints on a few essential braneworld parameters, not previously discussed in context of wormholes. In a nutshell, we have constructed a physically viable traversable wormhole without introducing any exotic matter at the throat, as necessary in the relativistic context. This is possible due to the higher dimensional contribution to gravity.

\section{DISCUSSIONS AND CONCLUSION}

In this work, we have successfully  obtained a physically acceptable solution describing a traversable wormhole
under the framework of braneworld gravity excluding exotic matter. Here we have calculated the shape function by
implementing  Kuchowicz type metric function as $g_{tt}$ in the field equations. The form of the adopted Kuchowicz potential to study wormhole is a well-behaved, analytic and non-singular function within the finite range of the radial coordinate. Though there is an inconsistency at $r \rightarrow \infty$ with the Kuchowicz metric function, but it provides the physically acceptable solutions for the wormhole providing a solution for the shape function which is well-behaved and satisfies all the essential criteria for being a shape function as prescribed by Morris and Thorne. Besides, it brings out the shape of the wormhole as well. Such a similar potential with identical features at r tending to infinity (Krori-Barua potential) has been used to construct wormhole solution\cite{IFI}. Various properties have been
studied in details to check the viability of our model. We have also estimated values of the model parameters
from junction conditions. We have actually used the Israel-Darmois junction conditions to obtain the values of the free constant parameters in our model. The junction conditions have been employed to obtain the values for the three free constant parameters in our model namely $X$, $Y$ and $\omega$ along with the two integration constants $C_1$ and $C_2$. For obtaining these values from the junction conditions using MAPLE, we have taken certain realistic values of the physical parameters involved in our model. There are two physical parameters for the brane, namely the brane tension and tidal charge. Our choice of the brane tension is for intermediate energy scales comparable to the present epoch of cosmic evolution~\cite{Sengupta} and the tidal charge which arises as a consequence of higher dimensional effect on any spherical symmetric matter distribution in its exterior spacetime, causes considerable modification to GR for very small values comparable to the order we have chosen. For the wormhole there is one physical parameter that we have chosen namely the throat radius, which is arbitrarily chosen and is not unique. It is usually chosen to be less than unity in most models. Varying it slightly does not impact our analysis significantly. One more physical parameter for the wormhole that we have assumed is the wormhole mass. Generally in the relativistic context it is taken to be very small as it is composed of unobserved exotic matter. So in order to minimize the quantity of the exotic matter used, it is chosen to be very small. But as in the RS brane scenario, we do not require any exotic matter to construct the wormhole, there is no need for choosing the mass to be extremely small to minimize the amount of composite matter, and we have taken a probable realistic value of 1.75 solar mass for our analysis.
As we shall note, we can find a satisfactory estimate of the bulk EoS parameter besides
drawing constraints on the brane tension. In this Section we are going to discuss some of the important results
that we have obtained from our present study.\\

1. \textit{Matter density}: By solving the conservation equation along with the EoS we have obtained the matter density
of the wormhole and its variation has been shown in Fig. 1. From the plot of the density with respect to the radial distance, it
can be confirmed that the matter near the throat is much denser than the boundary of the WH. This is in well agreement
with the tubular shape of wormhole as, on going radially outward from the throat the density decreases, i.e., matter get
distributed over the asymptotically flat spacetime.\\

2. \textit{Properties of Shape function}: For the construction of a wormhole the shape function plays an important role.
In order to obtain a physically acceptable traversable wormhole, the shape function must satisfy a number of conditions~\cite{MT1988}.
Here we discuss those conditions and corresponding behaviour for our model:\\

(i) First of all at the throat radius $r=r_0$, the shape function $b(r)$ must be equal to the
throat radius $r_0$ itself. From Fig. 2 one can observe that our model satisfies this condition.\\

(ii) The metric coefficient $(g_{rr})$ must be regular and well-behaved, i.e., $b(r)<r$ for $r>r_0$.
Again from Fig. 2 we have observed this condition is also satisfied for our present study
of wormhole.\\

(iii) In addition to the above, the nature of the plot for the shape function must be
such that if the curve is rotated with respect to the $b(r)$ axis, the $b(r)>0$ half for both positive
and negative $r$, must approximately resembles with the upper half of a wormhole. Fig. 2 tells us that it
does so.\\

3. \textit{Flaring-out condition}: To have a stable wormhole, it must obey the flaring-out condition at the throat in
order to ensure that there is no pinch-off making the wormhole non-traversable. The flaring-out condition states that
the first order derivative of the shape function with respect to $r$ at the throat must be less than $1$. Now, this
implies that the NEC must be violated. We see from Eq.~(\ref{eq17}) that it is not the case
for our matter distribution, but on the brane we are interested only about the effective matter distribution which as
seen from Eq.~(\ref{eq22}) and Fig. 3 violates NEC. This is a huge advantage of considering wormhole on the brane as
contrast to GR, because we find that there is no need to consider exotic matter but the effective matter description
due to local and non-local higher dimensional corrections enforce violation of NEC even with ordinary matter on the brane.\\

4. \textit{Geometrical nature and traversability condition}: By evaluating the radial component of the 4-acceleration to be positive, we infer that the wormhole has an "attractive" geometry which implies that the radial acceleration of the observer must be directed outward in order to avoid being pulled into the wormhole. Also, the radial tidal acceleration is computed at the throat and found to satisfy the tidal constraint which ensures travesarbility of the traveller without being pulled apart by the tidal forces. An additional constraint is obtained on the velocity of the traveller from the tangential tidal constraint. \\

5. \textit{Linear stability}: A linearized stability analysis is performed involving the sound speed and a constraint is obtained on the sound speed in terms of the parameters of the solution for the wormhole spacetime. The constraint is used to obtain the stable regions for the wormhole in Fig. (4). \\

6. \textit{Constraints on brane tension and estimate for $\omega$}:  We have found two very interesting features in connection with the brane tension, which is perhaps one of the most important regulating parameter in braneworld physics from our study:\\

(i) If the brane tension $\sigma$ is made negative, it turns out that the values of the parameters to be obtained
from the matching conditions become imaginary, which strongly rules out the possibility of a negative tension brane. Such a
brane is well-known to exhibit instabilities and have also been ruled out by previous investigations involving braneworld black holes~\cite{Marolf} and
fundamental braneworld physics~\cite{Charmousis}. \\

(ii) As the value of the brane tension $\sigma$ is lowered from the value $10^3\ TeV^4$ as considered for our analysis,
we can see that on lowering $\sigma$ upto $1 TeV^4$, the solution is found valid for wormhole construction, but on lowering
$\sigma$ further in $0.01-0.99\  TeV^4$ range, the essential conditions for wormhole formation are violated. Also, the condition
for violation of the NEC for effective matter description on the brane no longer holds true. Thus, we can claim that from
our present study that we have imposed a theoretical constraint on the brane tension such that always $\sigma >1$. This is
exactly in confirmation with similar constraint drawn on the brane tension from the perspective of previous theoretical
investigations~\cite{MaartensLR}. A possible explanation of this may arise from the fact that as we approach extremely
high energy scales then, $\sigma \rightarrow 0$, as discussed before. So, for energy scales corresponding to $\sigma < 1$,
quantum gravity effects must dominate and the effective description on the brane is no longer valid.

Further the value of the bulk EoS
parameter $\omega$ we obtain from the matching condition is $0.9157485630$ which is close to $+1$ and justifies the bulk space to be as $AdS_5$ (negative cosmological constant). \\

The RS braneworld gravity is inspired from the higher dimensional Superstring/M-theories and the extra terms
which appear in the field equation (2), may be physically interpreted as follows. The third term representing the
local correction basically encodes the quadratic correction terms to the stress energy tensor. This gives us a modified
effective fluid description on the brane, using ordinary matter. The final term represents the correction induced on the
brane due to the non-local bulk effect, and we find that the matter on the bulk makes the higher dimensional bulk spacetime
to be of Anti de Sitter (AdS) nature for successfully obtaining traversable wormhole solution. This is again in good
agreement with RS braneworld gravity.

We can talk about direct correspondence between the model we have used and the F(R) models described in \cite{NojiriR} on at
least two instances:. Firstly, the effect of the string inspired Gauss Bonnet (GB) term in modifying the field equations
have been described in \cite{NojiriR}. Such similar involvements in the braneworld context is important in establishing the
instability of a brane having a negative brane tension\cite{Charmousis}. Our theoretical model of the wormhole is also found
to be inconsistent for a negative tension brane. So, it can be said that the modification to the EFE has similar physical
impact as it rules out negative tension branes.

Secondly, in \cite{NojiriR}, for a particular form of F(R) gravity, the hierarchy problem is found to be solved. 
It resolves the issue as to why the gravitational force is weaker than the other fundamental interactions. It has been shown 
in \cite{Randall1} that on introducing a second brane the hierarchy problem can be solved too. The physical explanation is 
that the full strength of gravity can be realized only if we are in the higher dimensional bulk spacetime, but not on the (3+1)-brane
as gravity is free to propagate in the bulk, unlike the other standard model forces. Similarly, the explanation for dark
matter and galaxy rotation curves as provided by modified F(R) gravity discussed in ref. 32 can also be provided in a higher
dimensional RS braneworld setup\cite{SP}. The effect of higher dimensional bulk gravity or the gravitational effect of the 
matter on the second brane is found to explain the galaxy rotation curves. Also, the introduction of a viscosity term in the 
matter sector may provide a modified description equivalent to the effective matter description on the RS brane. We will 
consider such a modification in the matter sector identical to RS gravity in constructing traversable wormhole model in our future work.

\subsection{Observational status of wormholes}
Recently large number of attempts  have been made through observational evidences in favor of the idea of
braneworld~\cite{Abdujabbarov,Liddle,Keeton,Iorio,Visinelli,Vagnozzi} as well as on the possible ways for
detection of wormholes~\cite{Shaikh,Li,Ohgami,Tsukamoto,Nandi,Shaikh2}. Dai and Stojkovic~\cite{DS} have
suggested a possibility of observing a wormhole from the idea of non-conservation of individual fluxes in
the two different spacetimes connected by a wormhole, which may result in mutual effect on objects in close
approximate of either mouths of the wormhole due to one another. They have suggested the application of this
idea to study the possible effect on the orbit of stars near the black hole at our galactic centre, which may
possibly harbor a traversable wormhole. The possible production of micro-lensing effects by wormholes, resembling
gamma ray bursts, have been studied by Torres et al.~\cite{Torres} and successfully imposed a constraint on the
upper limit of the mass density of wormholes using the BATSE data. Doroshkevich et al.~\cite{D} have suggested a
possible technique of detection of wormholes by detecting the possible radiation pulse emitted from the other
side of a wormhole. The quasinormal ringing of black holes have been studied by Konoplya and Zhidenko~\cite{KZ},
who have investigated the scattering properties in the vicinity of a rotating traversable wormhole. A class of
non-symmetric wormholes is found to exhibit super radiance and symmetric wormholes may be easily distinguished
from black holes as their simultaneous ringing at various dominant multipoles are not identical. As per the
findings of Churilova et al.~\cite{C}, for wormholes with a variable redshift function and tidal force in the
radial direction, there exists long-lived quasinormal modes or quasi-resonances in the wormhole background.
Wielgus et al.~\cite{W} have constructed a reflection-asymmetric wormhole from two distinct RN spacetimes
with different mass and charge. Such a wormhole may be detected by the presence of unique, single or double
photon rings in its shadow due to reflection of a considerable amount of infalling radiation back to the spacetime
containing the source.

\vspace{0.5cm}

In many works on wormhole, the redshift function is simply assumed ad hoc to be a constant or an inversely proportional first order function of $r$ from a view point of mathematical simplicity or a well-known shape function known to satisfy all the properties of a traversable wormhole is assumed from the beginning. As a contrast to this, we have chosen a physically satisfactory well-behaved metric potential for spherically symmetric spacetime in the form of the Kuchowicz function, as the redshift function.

Two of the conditions for having a traversable wormhole are that the metric must be asymptotically flat and regular everywhere, and the Kuchowicz metric does not satisfy these conditions. The redshift function diverges as r tends to infinity. We are considering a model here for a static Lorentzian wormhole and as there is no time evolution or expansion of the wormhole throat, so the finiteness of the redshift function is physically significant only in the vicinity of the wormhole. It can be seen that for all finite values of r, the redshift function exhibits regularity and is capable of producing a well behaved and stable, traversable wormhole with the obtained shape function satisfying all the criteria as per the Morris-Thorne prescription, as has been discussed in the second point.
	
Using this setup, we have obtained naturally a solution for the shape function which agrees with all the essential criteria required to form a traversable wormhole (as prescribed by Morris and Thorne), just by naturally solving the field equations on the brane and also reproduces roughly the shape of the outer surface of a wormhole. We also find that there is no need for any exotic matter and the Null Energy Condition is violated automatically which we interpret to be the contribution of the  gravitational effect of higher dimensions, just like it provided the dark matter effect. The novelty of our work also lies in the fact that we have imposed constraints on the brane tension to be positive with a lower limit for successful description of a traversable wormhole which turns out to be in exact accordance with constraint obtained from other physical aspects.

The observational status of wormhole physics in general has been discussed briefly in subsection 3.1. From our model we can say that, if there is any particular signature in favour of higher dimensions from particle physics experiments in the near future, then the study of certain features of a higher dimensional brane universe like the brane tension or the nature of the higher dimensional bulk matter may be used to account for the possibility of physical existence of wormholes. It is evident from our model that in order to satisfy the physical characteristics of a wormhole-like object, some constraints are imposed upon the essential features of a higher dimensional universe. These features like the tension or the nature of bulk matter are in turn linked to properties of fundamental particles in particle physics. So, the validity of the constraints imposed by our wormhole model on these features may be used to rule out or favour our wormhole model.

In a nutshell, we have chosen such a gravitational framework for constructing wormhole, where the need for beyond the standard model exotic matter from a particle physics point of view is met up by the behaviour of gravity in the extra-dimensional scenario. So, obtaining the violation of the NEC without any exotic matter can be justified clearly. Also, the fact that the obtained shape function of the wormhole satisfies all the Morris-Thorne criteria and that the other physical features of the wormhole like the traversability and stability turn out to be satisfactory recommends that the idea of wormhole is more naturally compatible in the higher dimensional braneworld framework than in standard GR, as we do not have to put any exotic form of matter to stabilize the wormhole spacetime.

As a final comment we conclude that, our present analysis satisfies all the required conditions to ensure traversability in a static wormhole spacetime on the brane, without any exotic matter. Thus, we can justify the applicability of the Kuchowicz metric function to wormholes as well, in the presence of non-exotic matter on the braneworld. The result of our investigation is further strengthened by a probable estimation for $\omega$ and quantitative restriction imposed on $\sigma$ to be non-negative and greater than unity, which are in excellent agreement with the previously worked out models for braneworld black holes and fundamental braneworld physics. Though the existence of wormholes is yet to be directly confirmed, we may go on to state that if a wormhole is detected in the future, then our universe has a very high likelihood of being a 3-brane embedded in higher dimensions, due to our successful  construction of a wormhole on the brane in the absence of any kind of exotic matter, exactly satisfying the properties of the braneworld involving essential parameters like the brane tension. It is difficult to test the results experimentally at this moment. However, if a wormhole is detected in the near future and some constraint is obtained on its mass and throat radius, then there is a possibility that such a constraint restricts the brane tension to be positive as in our model and other brane gravity models\cite{Marolf,Charmousis} besides putting a lower constraint on the positive brane tension as obtained by us.

There has been an effective gravitational effect on the non-exotic matter present on the brane that has led to the violation of the NEC even in the absence of any exotic matter. The effect of viscous matter or effect of torsion and shear on such ordinary matter can be explored in future work, both in a relativistic as well as higher dimensional braneworld context. It would be interesting to study whether such a modification in the matter sector can allow the construction of traversable wormholes without exotic matter.
	
Moreover, there are other higher dimensional braneworld models like the DGP, or generalizations of the Randall-Sundrum scenario like the Sahni-Shtanov brane models. It would be interesting to explore the development of a traversable wormhole, particularly in the Sahni-Shtanov context, as such higher dimensional gravitational effects are well-known to admit non-singular bouncing solutions violating the Null Energy Condition in the cosmological context as well. Also, it would be of interest to explore the possibilities of construction of traversable wormholes in other physically attractive modified gravity scenarios which do not admit higher spacetime dimensions. It could then be realized whether the traversability of wormhole in the absence of exotic matter is a feature peculiar to extra spacetime dimensions or any gravitational framework admitting a class of modification to the relativistic Einstein-Hilbert action.

\section*{Acknowledgments}
The authors are thankful to the anonymous referees for their help in improving the manuscript substantially. MK and SR are thankful to the Inter-University Centre for Astronomy and Astrophysics (IUCAA), Pune, India for providing the Visiting Associateship under which a part of this work was carried out. RS is thankful to the Govt. of West Bengal for financial support through SVMCM scheme. SG is thankful to the Directorate of Legal Metrology under the Department of Consumer Affairs, West Bengal for their support.

~~~~~~~~~~~~~~~~~~~~~~~~~~~~~~~~~~~~~~~~~~~~~~~~~~~~~~~~~~~~~~~~~~~~~~~~~~~~~~~~~~~~~~~~~~~~~~~

\end{document}